\documentclass{aa}
\usepackage{graphicx}
\begin{document}

\title{Asymmetrical structure of ionization and kinematics in the Seyfert galaxy NGC 5033}

\author{E. Mediavilla \inst{1}, A. Guijarro \inst{2}, A. Castillo-Morales \inst{3}, J. Jim\'enez-Vicente \inst{3}, E. Florido \inst{3}, S. Arribas \inst{4}, B. Garc\'{\i}a-Lorenzo \inst{1}, E. Battaner \inst{3}}

\offprints{E. Mediavilla, emg@ll.iac.es}

\institute{Instituto de Astrof\'{\i}sica de Canarias, Tenerife, Spain
          \and Centro Astron\'omico Hispano Alem\'an, Almer\'{\i}a, Spain
          \and Dpto. F\'{\i}sica Te\'orica y del Cosmos, Universidad de Granada, Spain
          \and Space Telescope Science Institute, Baltimore,
          USA. Affiliated with the RSS Department of the European
          Space Agency. On leave from the IAC - CSIC}

\date{}
\authorrunning{Mediavilla et al.}
\titlerunning{Asymmetrical structure of ionization in NGC 5033}

\abstract{We present integral field spectroscopy of NGC 5033, a low luminosity Seyfert galaxy. The observations were made with INTEGRAL, a fiber based system operating at the WHT.  The intensity map of the H$\beta$ emission line represents a spiral or ring-like pattern of HII regions. On the contrary, the [OIII] intensity map morphology is markedly anisotropic. The strong morphological differences imply that the [OIII] emitters represent highly ionized gas illuminated by the central source. The [OIII] map morphology is compatible with a biconical structure of ionization induced by strong extinction in the galaxy disc that also obscures half of the spheroidal stellar bulge. We identify the spectrum corresponding to the Seyfert 1 nucleus from the presence of H$\beta$ broad emission lines. This spectrum is located in a region where strong extinction is expected but exhibits the bluest spectral energy distribution. The Seyfert 1 nucleus seems to be offcenter with respect to the stellar rotation center. This result has been also found in other Seyfert galaxies and interpreted in terms of a past merger. The offcentering could indicate the presence of nonsymmetric departures in the gravitational potential which could be fueling the active nucleus. The kinematics of the [OIII] emitters show important deviations at a kpc scale with respect to the stellar velocity field and show features related to the asymmetrical morphology of the high ionization region. 
}

\maketitle

\begin{figure*}
   \centering
   \caption{Intensity (arbitrary units) and velocity (kms$^{-1}$) maps for NGC 5033 (see text). The intensity maps correspond to the stellar continuum (integrated in the $\sim$4700\AA-5840\AA\ wavelength interval), the [OIII] $\lambda 5007$ and H$\beta$ emission. The star and the circle in the velocity maps represent, respectively, the location of the intensity peak in the stellar continuum and [OIII] intensity maps.}
   \label{fig:maps}
\end{figure*}

\section{Introduction}

According to observations, the narrow emission line regions in Seyferts and other active galaxies extend anisotropically over a significant fraction of the host galaxy. These Extended Narrow Line Regions (ENLR) trace the illumination of the gas by the ionizing continuum coming from the active nucleus that, in the framework of the unified models, should be collimated in two opposite cones along the axis of symmetry of a central obscuring torus. Sometimes, this biconical structure of ionization can be directly seen (e.g. NGC 5728, Arribas \& Mediavilla  1993, Wilson et al. 1993), but in most cases the scenario is complex and the shape of the ionization structure appears not clearly defined.
The shape and size of the ENLR depend not only on the relative inclination and widening of the radiation cones but also on the spatial distribution of the gas in the galaxy. Thus, the final appearance of the ENLR can be very amorphous if, as it seems to be the rule for high luminosity Seyfert nuclei, several gaseous systems of different spatial location and kinematics are ionized. In this case a kinematic analysis (as in NGC 1068, Arribas, Mediavilla \& Garc\'{\i}a-Lorenzo 1996) is needed to reveal the bipolar structure of ionization.  The study of the ENLR in the circumnuclear region of typical high luminosity Seyfert galaxies is also compromised since most of them are very distant. 

Nearby galaxies that host Low Luminosity Active Galactic Nuclei (LLAGNs) are potentially very interesting scenarios to study the ionization structure and, in general, to probe the unified scheme for AGNs. However the detection and mapping of the line emission in the circumnuclear region of LLAGNs is difficult. In these objects the emission lines are weak with respect to the continuum and filter imaging can give very doubtful results. A better observational option to study the ENLR is to obtain integral field spectroscopy from the circumnuclear region to directly measure the narrow line emission after properly fitting the continuum (see, for instance, the case of the LINER galaxy NGC 7331, Mediavilla et al. 1997; Battaner et al. 2003). With this aim, we present and discuss new observations of the low-luminosity Seyfert galaxy NGC 5033, taken with INTEGRAL, a fiber-based system to obtain integral field spectroscopy at the William Herschel Telescope. This galaxy has been alternatively classified as a Seyfert of type 1.5 (Ho et al. 1997c) or 1.9 (Veron-Cetty \& Veron 2001).   

NGC 5033 presents a very pronounced spiral structure and a nuclear bulge with a mass estimated to be $2\times10^{10}$M$_{\odot}$, to a distance of r = 1700 pc from the nucleus (Ford, Rubin \& Roberts 1971). The dynamics of nuclear activity of stars and gas in NGC 5033 have been studied. Bosma (1981) remarked that NGC 5033 has a giant neighbor within a short distance: NGC 5005, a Seyfert SAB(rs) galaxy. Using the Very Large Array, Thean et al. (1997) found neutral gas within a few kiloparsecs (1$\arcsec$ = 58 pc, D = 11.8 Mpc, H$_{\mathrm{o}}$ = 75 km s$^{-1}$ Mpc$^{-1}$) of the galactic center which appears to be unaffected by the active nucleus. From measurements of the optical emission lines (mainly H$\alpha$) and Westerbork HI-observations, van der Kruit \& Bosma (1978) did not detect any strong non-circular motion attributable to a bar. However, more recently Curran et al. (2001) with the 20m Onsala and 15m SEST telescopes detected a bar in HCN in NGC 5033. Ho \& Ulvestad (2001), using VLA for radio continuum observations at 6 cm and 20 cm emission, found that the morphology of the radio emission is predominantly that of a compact core slightly resolved surrounded by a fluffy envelope with a major-axis diameter of $\approx$ 0.9 kpc, roughly along the east-west direction, accompanied by elongated, jet-like features. Koratkar et al. (1995) and Terashima, Kunieda \& Misaki (1999) found that the X-ray emission is mostly or entirely nuclear ($\leq$ 500 pc) with a luminosity of $2.3\times10^{41}$ erg s$^{-1}$ in the 2-10 keV band. They also detected the Fe K$\alpha$ emission line at 6.4 keV. Recently, Kohno et al. (2003) have detected a perturbed distribution of CO in the central region of this galaxy with two bright peaks near the nucleus. 

In this paper we present integral field spectroscopy of the circumnuclear region of NGC 5033 to (i) identify its structure of ionization from the [OIII]$\lambda\lambda 4959,5007$ maps and (ii) study the kinematics of the ionized emitters from the comparison between the kinematics of stars and gas. 

\section{Observations and reduction}

NGC 5033 was observed with the fiber system INTEGRAL (Arribas et al. 1998) in combination with the fiber spectrograph WYFFOS (Bingham et al. 1994) at the William Herschel Telescope on 10 April 2001. The spectral resolution and coverage were 4.8\AA\ and 1445\AA\ and the central wavelength was 5003\AA\ . This galaxy was observed with the SB3 bundle of INTEGRAL. This bundle is formed of 135 fibers, each 2$\arcsec$.7 in diameter on the sky, and consists of a central rectangle of 30$\arcsec$$\times$34$\arcsec$ and an outer ring of 90$\arcsec$ of diameter. 3 exposures of 1800s and 4 of 600s were taken from NGC 5033. These frames were combined, in each case, to increase the S/N ratio and to reject the cosmic rays. Afterwards, the spectrum corresponding to each fiber was extracted, calibrated in wavelength, and corrected for throughput. The reduction steps were made using the INTEGRAL data reduction package running in IRAF. More details about the reduction procedure can be found in Arribas, Mediavilla \& Rasilla (1991). The average seeing was 1$\arcsec$.2, well under the size of a fiber face.

For several spectral features of interest we have determined a grid of points at the locations of the fibers at the focal plane and have interpolated a map of the selected feature in the region of the central rectangle  (see, e. g., Arribas et al. 1999). In Fig. 1 we present six maps obtained from the observations of NGC 5033. The stellar velocity field was obtained cross-correlating the spectra in the $\sim$5250\AA-6000\AA\ wavelength range. The average error of the cross-correlation is $50 \pm 17$ km/s. The continuum was integrated in the $\sim$4700\AA-5840\AA\ wavelength interval. To obtain the intensity and the velocity maps for the ionized gas we performed Gaussian fits to the [OIII]$\lambda5007$ and the H$\beta$ emission lines using DIPSO commands. To avoid the influence of strong stellar H$\beta$ absorption we considered in the fits an extra component representing this absorption feature. In the case of the [OIII] we obtain an average estimate for the velocity centroid uncertainties of $23\pm 10$ km/s.  The average uncertainty in the intensity is $\sim$ 10\%. In the case of H$\beta$ we obtain similar uncertainties in the spectra where appreciable emission was detected but we could not find emission in several fibers. Four of them close to the center of the field correspond to a region without H$\beta$ emission in Fig. 1.

\begin{figure}
   \centering
   \includegraphics[angle=0,width=8cm]{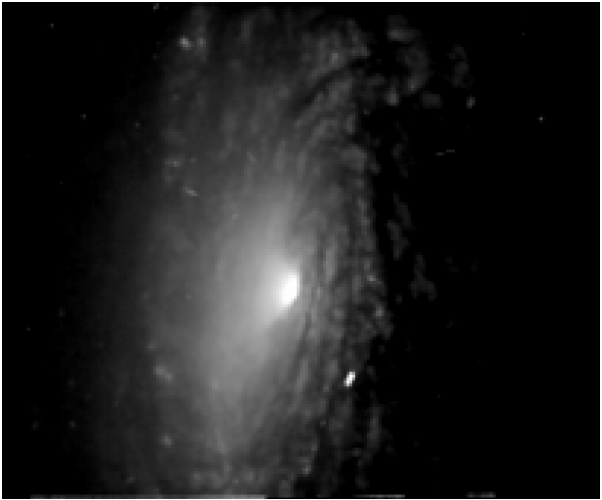}
   \caption{HST/PC (F606W) image of NGC 5033 (FOV: 36$\arcsec$x36$\arcsec$), obtained from the data archive at the Space Telescope Science Institute based on observations made with the NASA/ESA Hubble Space Telescope. East is to the left and North is up.}
   \label{fig:hst}
\end{figure}

\section{Results}

\subsection{Continuum and line emission maps}

The continuum intensity map looks very asymmetric with the emission peak (optical nucleus) apparently off-centered towards the west. To quantify this displacement we have fitted to the outermost closed isophote an ellipse (of semimajor axis $a$ = 13$\arcsec$.1)  obtaining that the center of the ellipse is  located $\sim$ 2$\arcsec$.1 towards the east of the optical emission peak. The orientation of the ellipse major axis (apparent photometric major axis) is PA $ 171^{\rm o}.1\pm 0^{\rm o}.3$. The continuum intensity map is morphologically identical to a map computed from the CaFe absorption feature (not shown here). This implies that it represents the distribution of stars and that the contribution from the Seyfert continuum is considerably smaller than the bulge emission. Clear evidence of extinction in the west side of the galaxy (closer to us) is found in an optical image from the HST archive (Fig. \ref{fig:hst}). Thus, we are likely observing half of the spheroidal stellar bulge of the galaxy, the other half being strongly obscured by extinction.

On the other hand, the peak of the [OIII] emission is close to the optical continuum emission peak but the [OIII] emission has a banana-like shape that does not follow the stellar continuum map (see Fig. 1). The H$\beta$ (narrow component) intensity map is very different from both the stellar and [OIII] intensity maps. In this map we can recognize several HII regions following a spiral or ring shaped pattern and a nuclear concentration displaced by about 2$\arcsec$ westward with respect to the [OIII] emission peak.

\subsection{Stellar and ionized gas velocity fields}

The stellar velocity field looks regular. We have fitted a kinematic ring model obtaining that the kinematic center is coincident within errors with the optical nucleus (offsets: 0$\arcsec$.03$\pm$ 0$\arcsec$.3 west and 0$\arcsec$.4$\pm$ 0$\arcsec$.4 south) and that the kinematic major axis is oriented along PA $ 170^{\rm o}\pm 6^{\rm o}$ parallel to the apparent photometric major axis (PA $ 171^{\rm o}.1\pm 0^{\rm o}.3$) of the galaxy (see Fig. 6). The [OIII] emission peak is $\sim$ 1$\arcsec$.5 westward from the kinematic center. The uncertainties in the determination of the kinematic parameters are estimated from the fit of the 2D distribution of velocities to the kinematic ring model and do not take into account that each measured velocity corresponds to the integration along the line of sight of a 3D stellar distribution that could be affected in a complex manner by selective extinction.

The ionized gas velocity field inferred from the Gaussian fitting to the [OIII] lines is related to the shape of the [OIII] emission. The  approaching (SE) pole morphology follows the bend of the [OIII] emission and the [OIII] emission peak appears strongly blueshifted. There are also noticeable departures from regular rotation in the region obscured by extinction ($\Delta\alpha <$ -8$\arcsec$). The ionized gas velocity field corresponding to H$\beta$ (see Fig. 1) is more regular than that of [OIII]. With the exception of the strongly blueshifted region near to the center, the H$\beta$ velocity field is compatible with a rotational pattern.

\section{Discussion}

\subsection{2D variation of the Spectral Energy Distribution (SED)}

\begin{figure}
   \centering
   \includegraphics[angle=270,width=8cm]{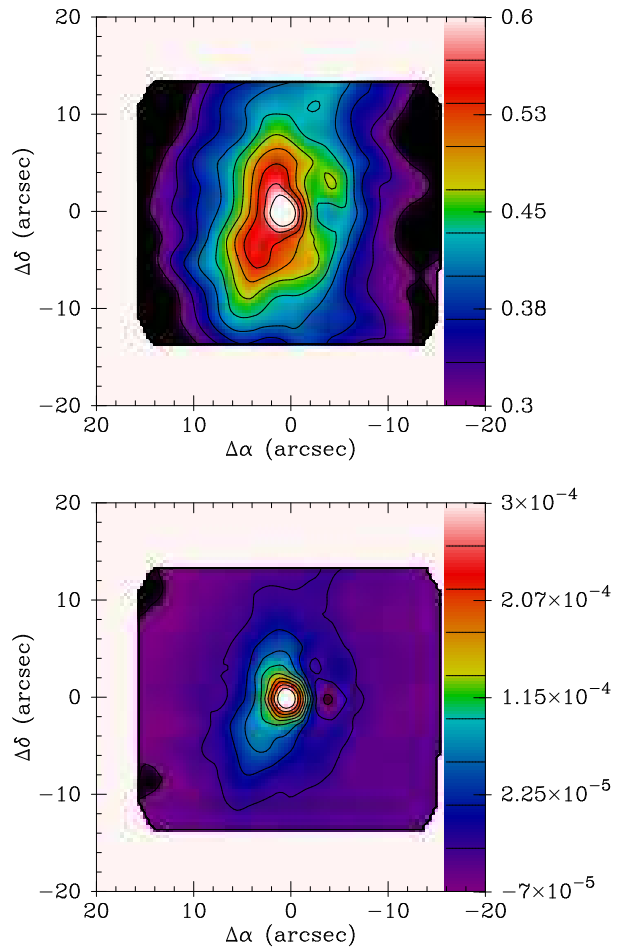}
   \caption{Top: 5965\AA/4700\AA\ color map (see text). Bottom: map of SED slopes (see text).}
   \label{fig:color}
\end{figure}

In Fig. \ref{fig:color} we present the quotient between a red continuum map obtained integrating the emission between 5930 and 6000 \AA\ and a blue continuum map integrated between 4600 and 4800 \AA. This color map shows the same morphology as the continuum emission (see Fig. \ref{fig:maps}) with the redder region coincident with the half non-obscured bulge. This correspondence cannot be induced by extinction and we should explore possible intrinsic variations in the SED in the observed region. To do this in a simple way we divide each observed spectrum, $f_\lambda$, by a template, $t_\lambda$, and fit the resulting function (in the wavelength regions free from emission lines) to a straight line:

$${f_\lambda\over t_\lambda}\sim a\lambda+b$$

As a template we take spectrum number 70 of high S/N ratio and without prominent emission lines. The fits are in general very good and we define rectified (drift-corrected) spectra as

$$f^*_\lambda={f_\lambda\over a\lambda+b}.$$

(The resemblance among the spectral absorption features of the rectified spectra is very noticeable; see Figs. \ref{fig:espectros1} and \ref{fig:espectros2}). 

In Fig. \ref{fig:color} (bottom) we show values of the parameter $a$ inferred from the fits. This map is similar to the color map (Fig. \ref{fig:color}, top) with the positive values of the slopes, $a$, following very closely the morphology of the half non-obscured stellar bulge. There is a singular point $\sim$ 4$\arcsec$ west from the nucleus, close to spectrum 71, with a very negative (blueward) value of the slope, $a$. This spectrum corresponds to a point where strong extinction is expected but that exhibits the bluest SED. This blue excess can be related to the emission of the unresolved active nucleus.

\subsection{Seyfert nucleus and BLR}

\begin{figure*}
   \centering
   \includegraphics[angle=0,width=18cm]{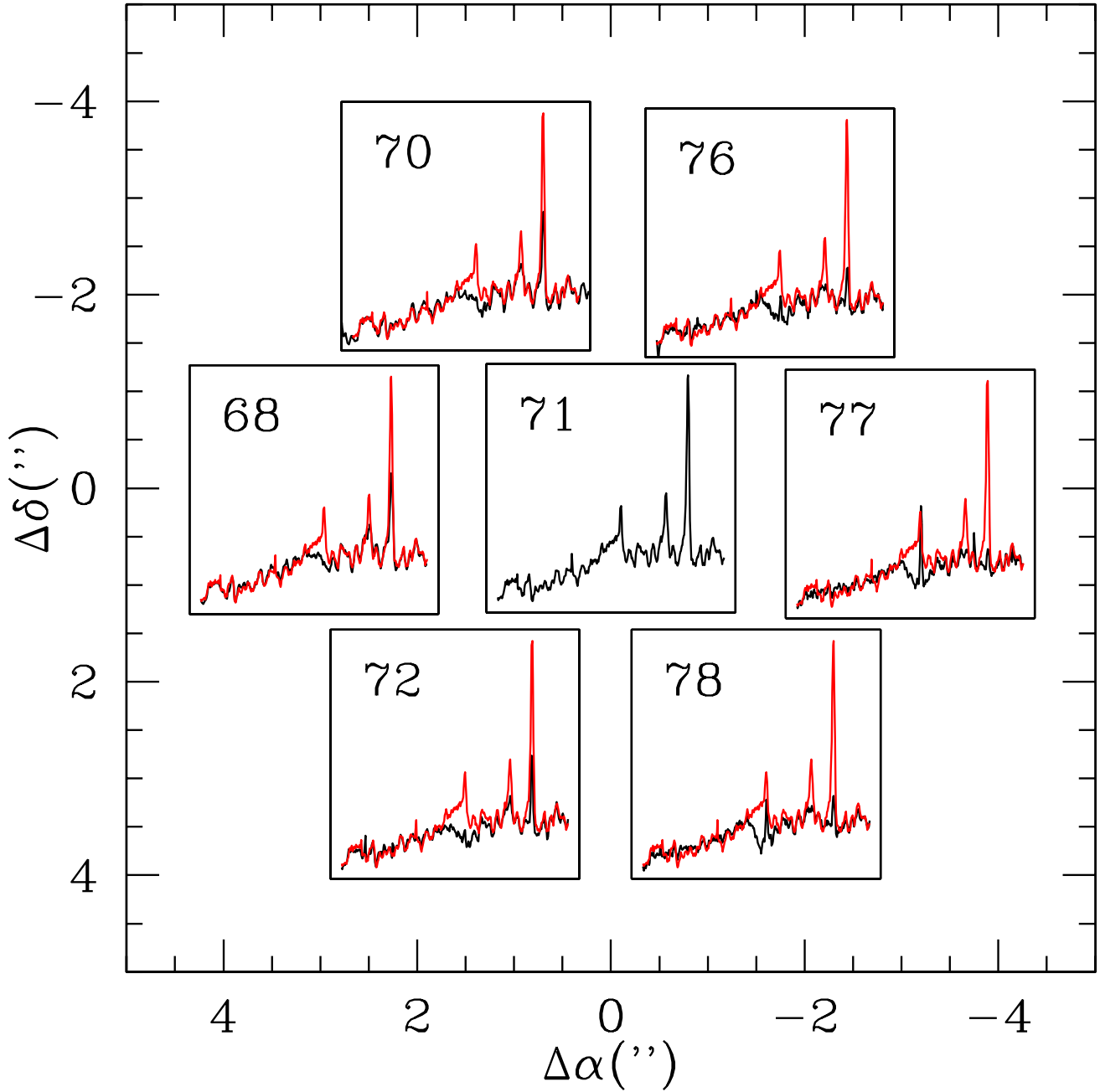}
   \caption{Comparison between spectrum 71 and adjacent spectra (the spectra have been drift-corrected, see text). Spectrum 71 is superimposed in red.}
   \label{fig:espectros1}
\end{figure*}

\begin{figure}
   \centering
   \includegraphics[angle=0,width=8cm]{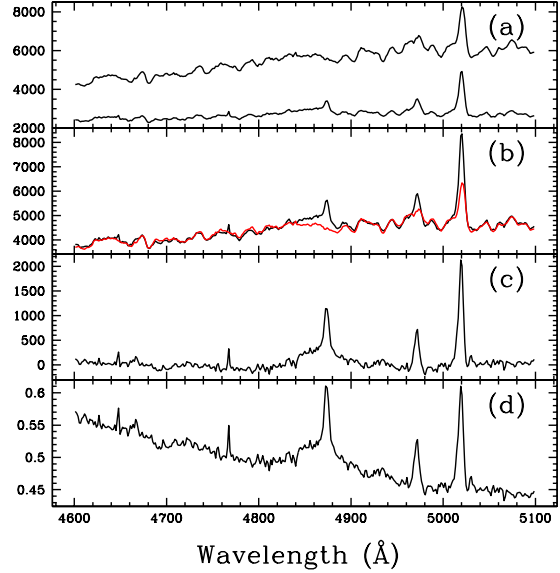}
   \caption{Comparison between spectra 68 (optical nucleus) and 71 (Seyfert nucleus). (a) Original spectra (68 top, 71 bottom). (b) Rectified spectra $f^*_\lambda(68)$ (red) and $f^*_\lambda(71)$ (black). Notice the good agreement among the absorption spectral features except in the region where the H$\beta$ broad emission lies in spectrum 71. (c) Difference between the rectified spectra, $f^*_\lambda(71)-f^*_\lambda(68)$.  (d) Quotient between the original spectra, $f_\lambda(71)/f_\lambda(68)$.}
   \label{fig:espectros2}
\end{figure}

Spectrum 71 also presents a significant broad H$\beta$ component. To show this explicitly we plot in Fig. \ref{fig:espectros1} the rectified spectra corresponding to fiber 71 and adjacent fibers.  In Fig. \ref{fig:espectros2} we compare the more detail spectrum 71 with the brightest one (68) which is almost coincident with the continuum peak. In Fig. \ref{fig:espectros2}c we represent the difference between both spectra showing clearly the broad H$\beta$ component in spectrum 71. Finally, in Fig. \ref{fig:espectros2}d we present the $f_\lambda(71)/f_\lambda(68)$ quotient where the blue excess of spectrum 71 (typical of AGN emission) is also clearly seen.

Thus, the continuum emission peak (at any wavelength in the observed range) falls in one fiber (number 68) and the H$\beta$ broad emission in another (number 71). This means that the Broad Line Region (BLR) and the continuum emission peak are not spatially coincident. Spectrum 71 is located at coordinates (-3$\arcsec$.2,0$\arcsec$.0). This implies that the Seyfert nucleus is, in the limits imposed by our spatial sampling, displaced $\sim$ 3$\arcsec$.2 westwards from the peak of the continuum emission. We can give as an upper limit for the uncertainty in the BLR location half the distance between the centers of two adjacent fibers, $\sim$ 1$\arcsec$.6, but this limit is very conservative for it would imply that for a 1$\arcsec$ seeing a noticeable part of the light coming from the BLR would be thrown to the adjacent fiber(s) and we do not find evidence of broad emission in any other fiber. It still might be possible that some broad component was hidden in the H$\beta$ absorption of spectrum 68. However, the comparison between the H$\beta$ absorption features of spectra 70, 68 and 72 do not support this possibility.

On the other hand, the (narrow) H$\beta$ emission peak is almost coincident with the location of the BLR. The broadening of the narrow H$\beta$ emission lines is practically the instrumental one in all the spectra except in spectrum 71. In this case, the measured broadening corresponds to an instrumentally corrected dispersion velocity of $\sigma \sim 140$ km/s. We found an [OIII]/H$\beta$ ratio of 4.3 for spectrum 71 that according to Whittle (1992) would correspond to a 1.8 type Seyfert.

Because of the offset between the BLR and the optical nucleus, previous spectra taken with long slits (Ho, Filippenko \& Sargent 1997a,b) of different width and aligned at different PA which lead to the classification of NGC 5033 as a variable low luminosity Seyfert 1.5-1.9 nucleus could not be compared with ours.

\subsection{Location of the galaxy center}

\begin{figure}
   \centering
   \includegraphics[angle=270,width=8cm]{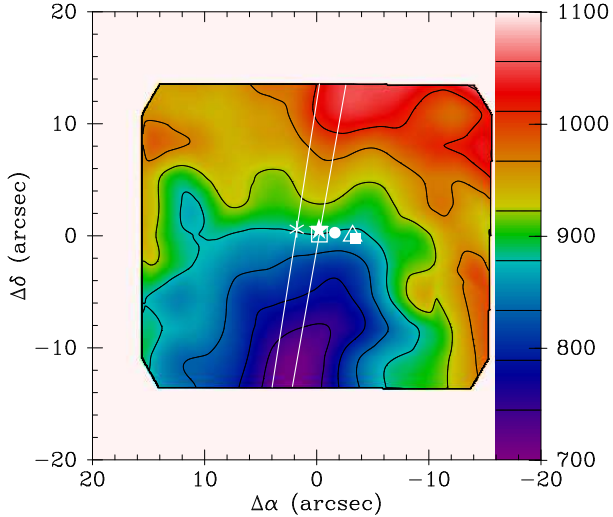}
   \caption{Photometric and kinematic centers. The eastern white line corresponds to the major axis of the ellipse fitted to the outermost isophote of the stellar continuum in Fig. 1 (apparent photometric major axis). The western white line is the kinematic axis inferred from a ring model fitting to the stellar velocity field. From left to right (east to west) the asterisk, star, square, circle, triangle and filled square represent the center of the ellipse fitted to the outermost continuum isophote, the stellar continuum emission peak, the kinematic center, the [OIII] emission peak, the BLR location and the H$\beta$ emission peak, respectively.}
   \label{fig:ejes}
\end{figure}

It is readily apparent that there are several non-coincident centers. They appear aligned almost perpendicular to the apparent photometric major axis (Fig. \ref{fig:ejes}). The offsets relative to the optical nucleus  (stellar emission peak) are: $\sim$ 2$\arcsec$.5 eastward (center of the ellipse fitted to the outermost continuum isophote), $\sim$ 1$\arcsec$.5 westward ([OIII] emission peak); and $\sim$ 3$\arcsec$.5 westward (H$\beta$ emission peak). The (narrow) H$\beta$ emission peak is almost coincident with the location of the BLR. The displacement of the apparent photometric major axis is likely produced by the strong extinction present in the west part of the disc of this galaxy. The optical nucleus also could have been affected by extinction, however it is coincident with the kinematic center (in principle less sensitive to extinction). This coincidence makes this location a good candidate for the galaxy center, leaving the BLR and the Seyfert nucleus (detected in a fiber different from the one corresponding to the optical nucleus, see \S4.2.) clearly offcentered by about 3$\arcsec$.

There are a number of galaxies in which the active nucleus is not located at the galaxy center, like M 31 (see, e.g., Kormendy \& Richstone 1995; del Burgo, Mediavilla \& Arribas 2000), NGC 1068 (Arribas, Mediavilla \& Garc\'{\i}a-Lorenzo 1996), NGC 3227 (Mediavilla \& Arribas 1993) and NGC 3516 (Arribas et al., 1997). In some cases the offcentering of the Seyfert nucleus has been related to a galaxy merger. The BLR would correspond to the Seyfert nucleus of a merged galaxy, orbiting the dynamic center that could constitute the center of the primary galaxy before the merging. These complex kinematics could indicate the presence of a nonsymmetric contribution to the gravitational potential. According to theoretical considerations (e.g., MacKenty et al. 1994, and references therein), this type of potential could redistribute the angular momentum of nonnuclear material, thereby bringing it closer to the galaxy center, exciting and/or fueling the nuclear activity. 

\subsection{Asymmetrical morphology of the high ionization region}

The strong differences in morphology between the H$\beta$ and [OIII] maps imply that the [OIII] map corresponds to highly ionized gas. The high ionization region is not symmetrical with respect to the nucleus but exhibits a banana-like shape. Elongated non symmetrical structures like this can be evidence of bi-conical (or bi-polar) emission (see e.g. NGC 1068, Arribas et al. 1996). In Fig. 7 we have interpreted the observed morphology in terms of a bi-conical emission taking as cone apex the BLR location. The west cone, strongly obscured by dust extinction, could be recognized in the two protrusions at coordinates $\sim$(-4$\arcsec$,-8$\arcsec$) and $\sim$(-7$\arcsec$,6$\arcsec$). Under this hypothesis, the apertures of the cones would be very wide and their axes would appear perpendicular to the disc galaxy plane. The cones would be hollow structures with a banana shape morphology. This could be caused by a lack of gas outside the galaxy plane more than by a feature of the ionizing flux.
All the morphological characteristics suggest that the asymmetrical morphology of ionization seen in the [OIII] map arises from the large scale distribution of obscuring material in this galaxy (notice, related to this, the discovery of a two-peaked distribution of CO aligned with the galaxy disc plane by Kohno et al. 2003). In Fig. 7 we include a FWHM map of the [OIII] emission lines which also present an asymmetrical aspect, perhaps related to the existence of several kinematic components integrated in the line of sight. 

 There are also traces of more collimated emission not perpendicular to the galaxy plane that could be constrained by some mechanism acting at a smaller spatial scale. These are one jet-like structure of low intensity at the West (PA $\sim 310\rm ^o$) and a broader and  stronger feature at the East (PA $\sim 110\rm ^o$).

\begin{figure}
   \centering
   \caption{Tentative ionization cones (white continuous lines) represented on (from top to bottom) the [OIII] line emission (arbitrary units), the stellar continuum (arbitrary units), the [OIII] velocity (km/s), and the [OIII] FWHM maps (\AA). The dotted dashed line is parallel to the photometric major axis.}
   \label{fig:conos}
\end{figure}

The stellar continuum morphology is in excellent agreement with the presence of an obscuring structure of the same characteristics as that inferred from the ionized gas distribution (see Fig. 7). This reinforces the idea that obscuration induced by the galaxy disc is giving rise to the observed ionization structure.

\subsection{Kinematics of the highly ionized gas}

After comparison of the stellar velocity map with the ones representing the ionized gas inferred from the [OIII] and H$\beta$ emission lines, it can be seen that the active nucleus is affecting the ionized gas kinematics in several ways: (i) a region near the BLR location appears blueshifted in both the [OIII] and H$\beta$ maps (nuclear blueshifts have been found in the narrow emission lines of many Seyfert galaxies, see Whittle 1992 and references therein), (ii) the kinematic major axis corresponding to the [OIII] map is displaced with respect to that of the stars following the banana-like shape of the [OIII] intensity map (notice the strong displacement of the blue pole of the [OIII] velocity field toward the east with respect to that of the stars), (iii) there are departures from regular rotation in the west region of the [OIII] velocity map (not present in the H$\beta$ velocity field). Features (i) and (ii) may be related to the biconical geometry (see Fig. 7): the nuclear blueshifted region starts from the BLR location and extends along the NE cone border and the kinematic major axis in the SE region is also aligned with the SE cone border.

To study the non rotational kinematics we have computed a difference velocity map subtracting the stellar velocity map from the [OIII] velocity map. The resulting difference map (Fig. 8) is very irregular and difficult to interpret further than the previous comments. It is clear that the highly ionized gas is following peculiar kinematics neither related to axisymmetric motions nor ruled by the dynamic influence of a bar, but it is difficult to recognize a velocity pattern in this map. The role of activity is not limited to the illumination of the gas but it exerts a significant influence on the dynamics of the high ionization gas at kpc scale.

Regarding the jet-like feature likely representing more collimated emission toward the NW, it appears strongly redshifted in the difference velocity map.

\begin{figure}
   \centering
   \includegraphics[angle=270,width=8cm]{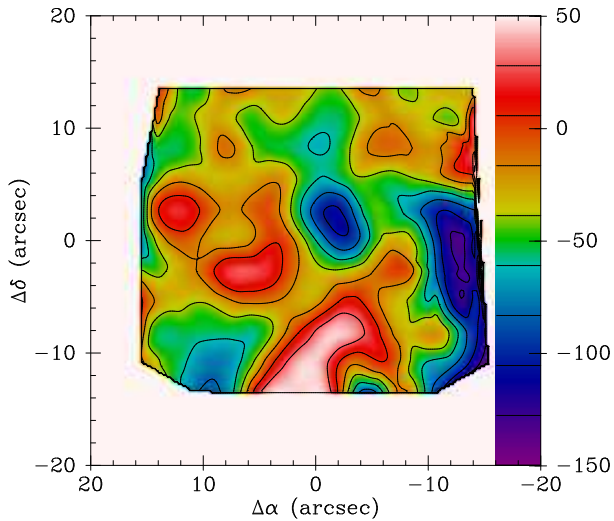}
   \caption{Residual velocity field obtained subtracting the stellar velocity map from the [OIII] velocity map.}
   \label{fig:dif}
\end{figure}

\section{Conclusions}

1 - The location of the Seyfert 1 nucleus of the galaxy that we infer from the location of the broad H$\beta$ emission is $\sim$ 3$\arcsec$.2 westwards of the continuum emission peak. The stellar rotation center seems to be located close to the continuum emission peak. According to the analysis of the observations presented in \S4.3., both centers (galaxy mass centroid and active nucleus) cannot be reconciled. The offcentering of the Seyfert nucleus found in other active galaxies has been considered evidence of a past merger event possibly related to the origin or to the fueling of the activity.

2 - The [OIII] emission (associated with highly ionized gas) depicts an asymmetrical morphology which could be interpreted in terms of a bi-conical ionization structure with its apex in the Seyfert 1 nucleus. The cones could be hollow structures of wide aperture and projected axis perpendicular to the galaxy major axis. The asymmetrical shape of the high ionization region seems to be related to obscuring material in the galaxy disc that also induces the peculiar morphology of the stellar continuum emission (with half of the stellar bulge obscured).

\begin{acknowledgements}

The Isaac Newton Group of Telescopes (ING) operates the 4.2m William
Herschel Telescope on behalf of the Particle Physics and Astronomy 
Research Council (PPARC) of the United Kingdom and the Netherlands
Organization for Scientific Research (NWO) of the Netherlands. The ING
is located at the Roque de Los Muchachos Observatory, La Palma, Spain. 

Thanks to the anonymous referee for valuable comments and suggestions that have contributted to improve the final version of the paper.

This paper has been supported by the Euro3D RTN ``Promoting Integral Field Spectroscopy in Europe'' (HPRN-CT-2002-00305) by the ``Plan Andaluz de Investigacion'' (FQM-108) and by the ``Secretar\'{\i}a de Estado de Pol\'{\i}tica
Cient\'{\i}fica y Tecnol\'ogica'' (AYA2000-2046-C02-01). J.J.V. acknowledges support from the Consejer\'{\i}a de Educaci\'on y Ciencia de la Junta de Andaluc\'{\i}a, Spain. 

Some of the data presented in this paper were obtained from the 
Multimission Archive at the Space Telescope Science Institute 
(MAST). STScI is operated by the Association of Universities for
Research in Astronomy, Inc., under NASA contract NAS5-26555.

\end{acknowledgements}

\end{document}